\documentclass[%
reprint,
superscriptaddress,
amsmath,amssymb,
aps,
prl,
]{revtex4-1}
\usepackage[utf8]{inputenc}
\usepackage{graphicx}
\usepackage{dcolumn}
\usepackage{mathrsfs}
\usepackage{bm}

\begin{document}


\title{Evidence for the Single-Site Quadrupolar Kondo Effect in the Dilute non-Kramers System Y$_{1-x}$Pr$_x$Ir$_2$Zn$_{20}$}

\author{T. Yanagisawa}
 \affiliation{Department of Physics, Hokkaido University, Sapporo 060-0810, Japan}
\author{H. Hidaka}
 \affiliation{Department of Physics, Hokkaido University, Sapporo 060-0810, Japan}
\author{H. Amitsuka}
 \affiliation{Department of Physics, Hokkaido University, Sapporo 060-0810, Japan}
\author{S. Zherlitsyn}
 \affiliation{Hochfeld-Magnetlabor Dresden (HLD-EMFL), Helmholtz-Zentrum Dresden-Rossendorf, 01328 Dresden, Germany}
\author{J. Wosnitza}
 \affiliation{Hochfeld-Magnetlabor Dresden (HLD-EMFL), Helmholtz-Zentrum Dresden-Rossendorf, 01328 Dresden, Germany}
 \affiliation{Institut für Festkörper- und Materialphysik, TU Dresden, 01062 Dresden, Germany}
\author{Y. Yamane}
\affiliation{
 Graduate School of Advanced Sciences of Matter, Hiroshima University, Higashi-Hiroshima 739-8530, Japan
}
\author{T. Onimaru}
\affiliation{
 Graduate School of Advanced Sciences of Matter, Hiroshima University, Higashi-Hiroshima 739-8530, Japan
}

\date{\today}

\begin{abstract}
Acoustic signatures of the single-site quadrupolar Kondo effect in Y$_{0.966}$Pr$_{0.034}$Ir$_2$Zn$_{20}$ are presented. The elastic constant ($C_{11}-C_{12}$)/2, corresponding to the $\Gamma_3$(E)-symmetry electric-quadrupolar response, reveals a logarithmic temperature dependence of the quadrupolar susceptibility in the 
low-magnetic-field region below $\sim$0.3 K. Furthermore, the Curie-type divergence of the elastic constant down to $\sim$1 K indicates that the Pr ions in this diluted system have a 
non-Kramers ground-state doublet. These observations evidence the single-site 
quadrupolar Kondo effect, as previously suggested based on specific-heat and 
electrical resistivity data.
\end{abstract}
\pacs{Valid PACS appear here}
\maketitle

The Kondo effect~\cite{1Kondo64}, in a broad sense, appears ubiquitously as a phenomenon not only in dilute magnetic alloys or dense Kondo systems but also in various fields of physics~\cite{2Coleman15,3Coleman07}. In particular, heavy-electron physics has been intensively studied based on the Fermi-liquid theory originally introduced by Landau~\cite{4Landau56}, where a local Fermi-liquid state is generally realized due to the Kondo effect driven by local magnetic impurity. An unconventional Kondo effect with concomitant non-Fermi-liquid (NFL) behavior of the physical quantities has been discussed since the 1990s (see [\onlinecite{5Maple91}] for a review) and some of the compounds that evidence NFL behavior have been recognized to realize a new type of Kondo phenomena, which is caused by over-screening of the local multipolar moment via two or more channels, the so-called multi-channel Kondo effect~\cite{6Nozieres80}. The two-channel version of the multi-channel Kondo effect~\cite{Sacramento91} is the so-called quadrupolar Kondo effect (QKE), which was theoretically proposed by Cox~\cite{Cox88, 7Cox98}.

The quadrupolar Kondo model, which is invoked by a well-separated non-Kramers doublet $\Gamma_3$ (E) ground state in cubic symmetry, predicts several NFL peculiarities at very low temperatures: such as a temperature dependence of the normalized electrical resistivity $\rho/\rho_0 \propto 1\pm\sqrt{T}$~\cite{Affleck93}; the presence of a fractional residual entropy $S=\frac{1}{2}R\ln2$ at absolute zero; and a logarithmic temperature dependence of the specific heat divided by temperature, $C/T$, and also, of the quadrupolar susceptibility $\chi_{\Gamma}$ well below the characteristic temperature~\cite{Kusunose16}. In particular, the appearance of $-\ln{T}$ behavior in the $\Gamma_3$ quadrupolar susceptibility in the Cubic system can be observed as $+\ln{T}$ decreasing (softening) of the elastic constant $(C_{11}-C_{12})/2$ by means of ultrasound.

Many experimental investigations have been conducted to test this promising theoretical scenario, especially for uranium-based candidate materials, which were considered to have quadrupolar degrees of freedom and obey single-site NFL behavior, such as cubic Y$_{1-x}$U$_x$Pd$_3$ and Th$_{1-x}$U$_x$Be$_{13}$, as well as tetragonal Th$_{1-x}$U$_x$Ru$_2$Si$_2$~\cite{8Seaman91, 9Aliev94, 10Amitsuka94}. However, it is still elusive whether the quadrupolar degrees of freedom are involved in the NFL behavior found in these compounds, because no convincing evidence has been found with respect to the main cause of these exotic phenomena, {\it i.e.,} the local quadrupolar response. Due to the duality of the 5$f$ electrons (being partially localized as well as itinerant)  and experimental difficulties, a non-magnetic doublet ground state of the U 5$f$-electron system is possibly inappropriate to corroborate the QKE.

\begin{figure}[b]
\includegraphics[width=0.75\linewidth]{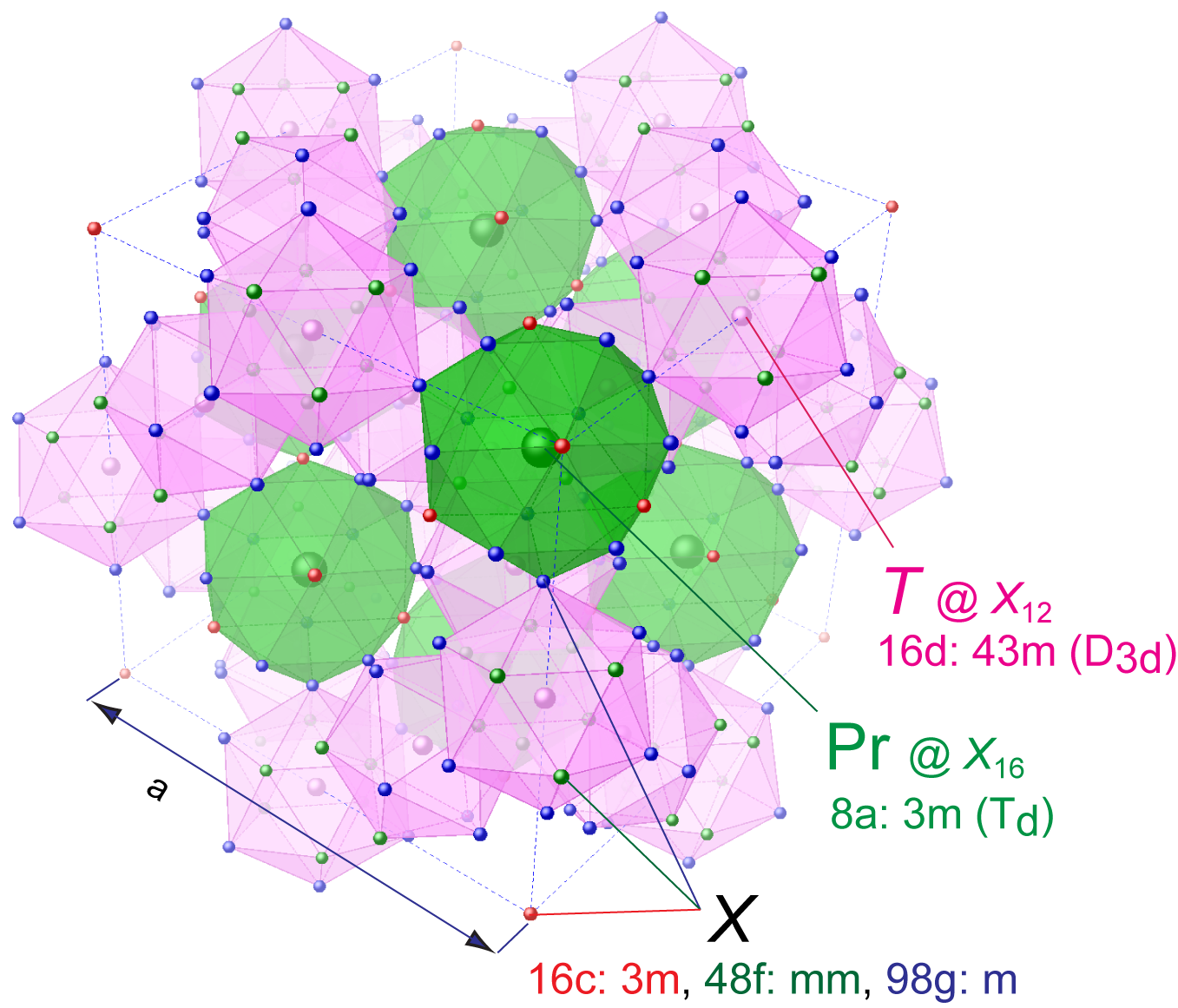}
\caption{\label{fig:fig1} Crystal structure of Pr$T_2X_{20}$ ($T$ = transition metal, $X$ = Al, Zn, and Cd). The Wycoff position and point-group symmetry of the atoms are also represented below each annotation.}
\end{figure}

Cubic Pr-based non-Kramers doublet systems, such as PrInAg$_2$, PrPb$_3$, and PrMg$_3$, have since been included as candidate materials for the QKE of the Pr$^{3+}$ (4$f^2$, $J=4$) state, which has paved the way to corroborate Cox's quadrupolar-Kondo scenario of the U$^{4+}$ (5$f^2$, $J=4$) state~\cite{12Yatskar96, 13Bucher72, Onimaru05, Kawae06, 14Tanida06}. A new family of cubic compounds, Pr$T_2X_{20}$ ($T$ = transition metal, $X$ = Al, Zn, and Cd), has recently attracted considerable attention~\cite{15Onimaru16}. Figure 1 shows the crystal structure of the Pr$T_2X_{20}$ compounds, which have the CeCr$_2$Al$_{20}$-type structure (Fd$\bar{3}$m, O$_h^7$, No. 227), where a Pr guest ion (on the site 8a with T$_{\rm d}$ symmetry) is encapsulated in a highly symmetric atomic cage consisting of $X_{16}$, which forms a diamond structure in the unit cell~\cite{16Sato12,17Iwasa13}.

 In this cage-structured-compound family, the Pr tends to have a $\Gamma_3$(E) doublet crystalline electric field (CEF) ground state. In particular, the Pr dilute-limit Y$_{1-x}$Pr$_x$Ir$_2$Zn$_{20}$ system is studied systematically to investigate a possible single-site quadrupolar Kondo-state~\cite{18Yamane18,19Yamane18,20Yamane18}. Y$_{1-x}$Pr$_x$Ir$_2$Zn$_{20}$ displays particular rich physics, ranging from a localized quadrupolar order of the $\Gamma_3$(E) non-Kramers doublet ground-state and superconductivity for $x$ = 1~\cite{21Onimaru10, 22Ishii11, 23Iwasa17} to possible single-site quadrupolar Kondo behavior for $x \to 0$, which appears as NFL behavior in the specific heat and resistivity. Here, we present data from ultrasonic measurements for the dilute-limit Y$_{0.966}$Pr$_{0.034}$Ir$_2$Zn$_{20}$ ($x=$ 0.034). We find acoustic signatures of the $\Gamma_3$(E) ground-state doublet, and a logarithmic temperature dependence of the $\Gamma_3$(E)-symmetry quadrupolar susceptibility, both of which give unambiguous evidences for the single-site quadrupolar Kondo effect.

Single crystals of Y$_{1-x}$Pr$_x$Ir$_2$Zn$_{20}$ were grown by Zn-self-flux method with pre-arc-melting alloys of Y, Pr, and Ir as described in previous papers~\cite{18Yamane18, 20Yamane18}. The Pr composition $x=0.034$ of the present sample is confirmed by the CEF analysis of the magnetization data, which is the same procedure used in the previous specific heat experiments~\cite{19Yamane18}. The sample dimensions of the rectangular parallelepiped are $2.954\times2.771\times2.440$ mm$^3$ for [110]-[1$\bar{1}$0]-[001]. Ultrasound is generated and detected by a pair of LiNbO$_3$ transducers with a thickness of 100 $\mu$m, which were bonded on the sample surfaces with room-temperature-vulcanizing silicone. The quadrupolar responses can be observed as sound-velocity change by the phase-comparative method. The transverse sound velocity $v$ with propagation of $k \parallel$ [110] and polarization $u \parallel$ [1$\bar{1}$0] is converted to the elastic constant $C_v$ = $(C_{11}-C_{12})/2$ (J/m$^3$) by using the formula $C_v$ = $\rho v^2$. Here, $\rho = 8.277$ (g/cm$^3$) is the density of the $x = 0.034$ sample with the lattice constant $a =  14.1969(1)$\AA. The low-temperature measurements were performed using a top-loading dilution refrigerator.

\begin{figure}[t]
\includegraphics[width=0.8\linewidth]{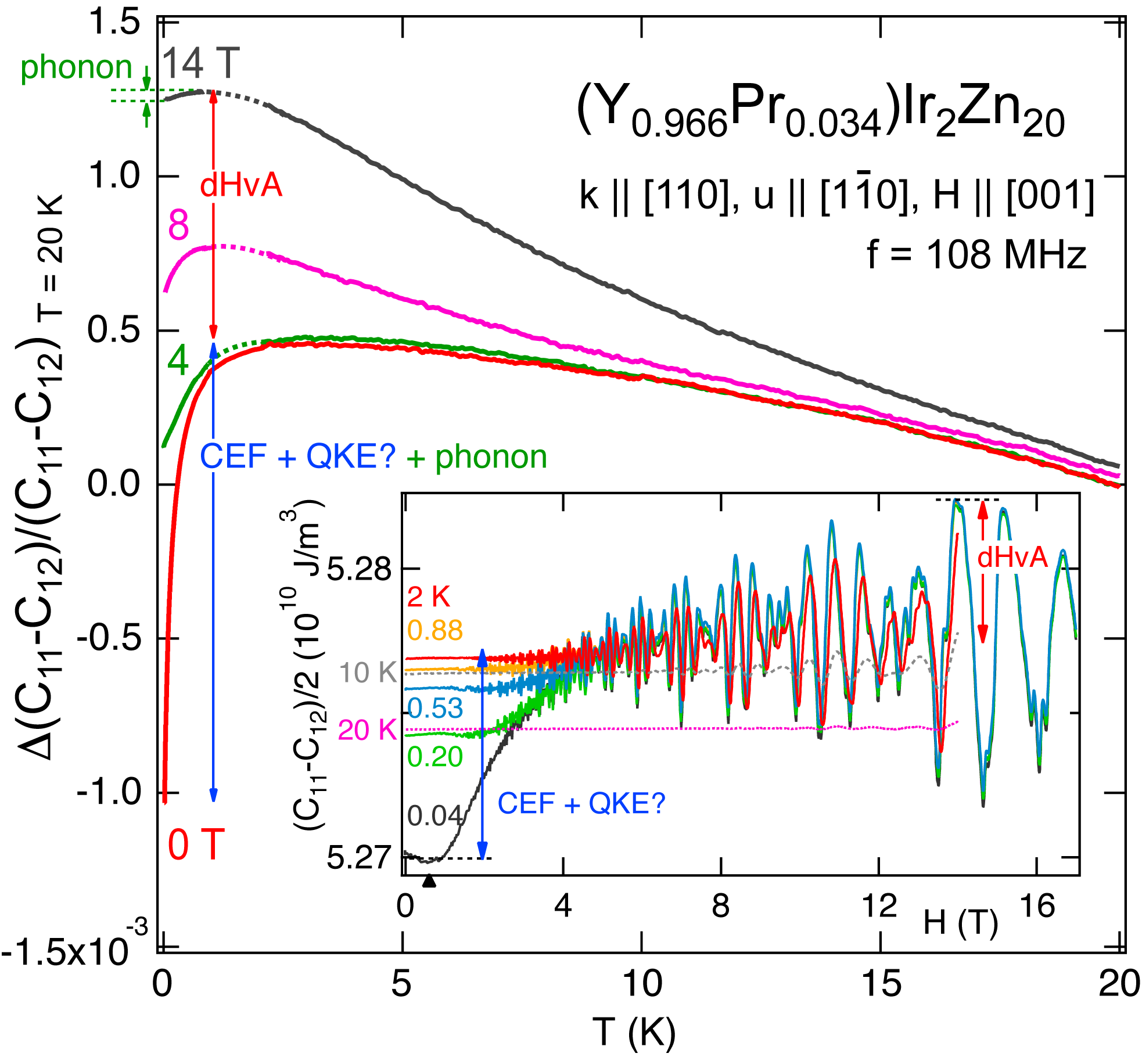}
\caption{\label{fig:fig2}  Relative change of the elastic constant $(C_{11}-C_{12})/2$, as a function of temperature for various magnetic fields up to 14 T for $H \parallel$ [001] of Y$_{0.966}$Pr$_{0.034}$Ir$_2$Zn$_{20}$. The inset shows $(C_{11}-C_{12})/2$ as a function of magnetic field at various fixed temperatures between 0.04 and 20 K.}
\end{figure}

The temperature dependence of the relative change of the elastic constant $(C_{11}-C_{12})/2$ is displayed in Fig. 2. At 0 T, we found Curie-like divergence (softening) of $\sim0.18\%$ in $(C_{11}-C_{12})/2$ below 2 K down to 0.04 K. The softening is reduced by an external magnetic field. At 14 T, there remains a small magnetically robust softening of $\sim0.02\%$, which might be caused by quantum tunneling of Zn atoms at the site 16c and/or Pr and Y atoms at the site 8a, so called, off-center tunneling (for details, see the supplementary materials)~\cite{24Goto04, 25Wakiya16}.

The temperature and magnetic-field dependence of the elastic constant corresponding to the point group symmetry $\Gamma$ in local 4$f$-electron system can be described as a sum of three components of the elastic moduli, $C_\Gamma (T,H) = C_\Gamma^0(T) + C_\Gamma^{4f} (T,H) + C_\Gamma^{\rm dHvA} (T,H)$. Here, $C_\Gamma^0(T)$ is the temperature dependent (magnetic-field-independent) phonon background with the contribution from the off-center tunneling. $C_\Gamma^{4f} (T,H)$ is the contribution from the 4$f$ electrons, mainly due to CEF effects, which can be described here by the quadrupolar susceptibility of the Pr ion and leads to the Curie-type softening~\cite{26Luthi}. $C_\Gamma^{\rm dHvA} (T,H)$ is the de Haas-van Alphen (dHvA) effect which leads to quantum oscillations in high-magnetic fields with frequencies proportional to $1/H$~\cite{KataokaGoto}. Since $C_\Gamma^0(T)$ is usually unaffected by magnetic field, we can simply estimate the latter two components in above equation by comparing the $T$ and $H$ dependence of the elastic constant $C_{\Gamma 3}(T,H)= (C_{11}-C_{12})/2$.

The inset of Fig. 2 shows the magnetic-field dependence of the $(C_{11}-C_{12})/2$, where acoustic dHvA oscillations with frequencies of 148, 185, 332, and 295 T (decreasing order of oscillation amplitude) are clearly observed below 20 K. Detailed analyses of these dHvA data are subject of a forthcoming publication, however, this observation clearly proves the high quality of the present single crystal.

In the main panel of Fig. 2, the red, green, and blue double-headed arrows indicate the amount of change of $C_\Gamma^{\rm dHvA}$ for 14 T, estimated from the field dependence, phonon contribution $C_\Gamma^0(T)$ below 1 K, and the softening originated in $C_\Gamma^0(T) + C_\Gamma^{4f} (T,H)$ for 0 T below 1 K, respectively. Here, it is obvious that the major part of the low-temperature softening in the present compound can be interpreted by $C_\Gamma^{4f} (T,H)$, which is a combination of CEF effects, nuclear effects, and a possible contribution from the QKE. Note that the temperature dependence of the dHvA oscillation amplitude $C_\Gamma^{\rm dHvA} (T)$ below 1 K is negligible compare to the other contributions because they only change within $\Delta C_v/C_v \sim2\times 10^{-5}$ for 3 T and $\sim1\times10^{-4}$ for 13.5 T.
 Based on the presence of the Curie-like softening ($\propto T^{-1}$) in $(C_{11}-C_{12})/2$ at low magnetic fields, it can be concluded that the Pr ions in the present dilute system also have a non-Kramers $\Gamma_3$(E) CEF ground-state doublet, which is crucial for QKE.
The amount of change in ($C_{11}-C_{12}$)/2 in the present Y-rich system, $\Delta C_v/C_v \sim0.18$\%, is larger than the expected value 0.013\% for $x$ = 0.034, which is simply estimated by multiplying the volume fraction of the Pr concentration to the softening of 0.39\% observed at the antiferroquadrupolar ordering temperature $T_{\rm Q} \sim0.11$ K in the non-diluted compound (PrIr$_2$Zn$_{20}$)~\cite{27Onimaru11, 22Ishii11}. A possible reason for this discrepancy is the absence of the inter-site quadrupolar-quadrupolar interactions in the present diluted system (as described later), since the amount of softening of the elastic constant will depend on the tradeoff between the enhancement of the softening due to the volume fraction of Pr-ion, and the reduction of the softening due to the antiferro-type of the inter-site quadrupolar-quadrupolar interactions. Further investigations of this system toward a wide range of Pr concentrations are needed to check a possible systematic change of the softening and the inter-site quadrupolar interaction.

\begin{figure}[t]
\includegraphics[width=0.85\linewidth]{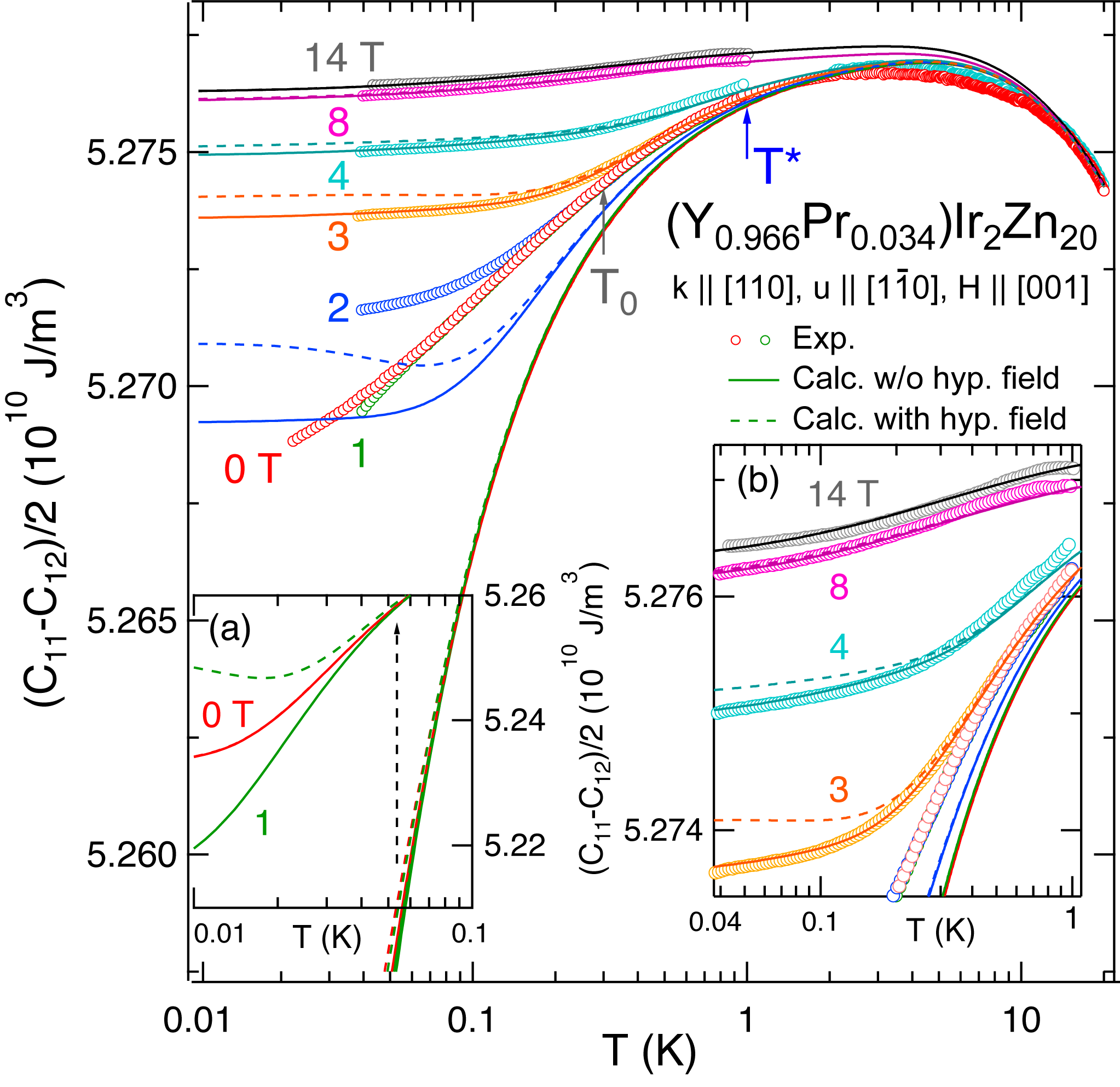}
\caption{\label{fig:fig3} Elastic constant $(C_{11}-C_{12})/2$ (open symbols) of Y$_{0.966}$Pr$_{0.034}$Ir$_2$Zn$_{20}$ as a function of temperature at various magnetic fields $H \parallel$ [001]. Dashed and solid curves indicate the calculated quadrupolar susceptibility with and without the $^{141}$Pr's nuclear dipolar contribution, respectively (see text). The insets show zooms of (a) the calculations below 0.1 K for 0 and 1 T, and (b) the data above 3 T.}
\end{figure}

In Fig. 3, the elastic constant $(C_{11}-C_{12})/2$ is represented as a function of temperature in logarithmic scale at various magnetic fields. Here, the contributions from dHvA efffect below 1 K, and are estimated from Fig. 2 which is already subtracted as a constant background. Dotted and solid curves are the calculated quadrupolar susceptibility based on the local 4$f$-electron state of Pr$^{3+}$ with and without the $^{141}$Pr's nuclear dipolar contribution, respectively. From the analysis, we notice that the local 4$f$-electron model well reproduces the data in 3 T and above, however not below 3 T and $T < T^* \sim1$ K, where the expected softening of $(C_{11}-C_{12})/2$ is strongly suppressed and also exhibits a logarithmic temperature dependence below $T_0 \sim0.3$ K. Here, the characteristic temperature $T_0$ is estimated from previous specific-heat measurements~\cite{20Yamane18}. Based on the similarity of the logarithmic temperature dependence found in the specific heat and elastic constant, we may assume that there is a strong correlation between the NFL behavior and the quadrupolar fluctuations in the present system.

Since we found no obvious frequency dependence of the ultrasound velocity in the range from 65 to 230 MHz (see supplementary material) and also no considerable ultrasonic attenuation at around $T_0$, a dynamical quadrupolar effect can be neglected, at least in the present frequency range. Thus, we analyze the elastic softening of $(C_{11}-C_{12})/2$ using the following formula describing the static quadrupolar susceptibility: $C_{\Gamma3}^{4f}(T, H) = C_{\Gamma3}^0(T)-Ng_{\Gamma3}^2\chi_{\Gamma3}(T, H)/\{1-g'_{\Gamma3}\chi_{\Gamma3}(T, H)\}$ (see the supplementary material and Ref.~\cite{28Yanagisawa18}). Here, $C_{\Gamma3}^0(T)$ is a magnetically independent background term, $N = 0.095\times10^{27} $ m$^{-3}$ is the number of Pr ions per unit volume in the $x$ = 0.034 sample estimated from the lattice constant at room temperature. $g_{\Gamma3} = 19.0$ K is the coupling constant of the quadrupole-strain interaction, which is obtained from the best fit for the data above 3 T, where the non-magnetic Kramers doublet obeys Zeeman splitting and mixing with the wave function from the first excited $\Gamma_4$(T$_1$) triplet CEF state (at $\sim25.3$ K in zero magnetic field). The coupling constant of inter-site quadrupolar interaction is set to $g'_{\Gamma3}=0$. This negligible inter-site quadrupolar interaction, as obtained from our data for the diluted system, allows to discuss the following single-site quadrupolar effect.

The $\Gamma_3$(E)-symmetry quadrupolar susceptibility $\chi_{\Gamma3}(T, H)$ for $H \parallel$ [001] is calculated by using the quadrupolar operator $O_2^2 = J_x^2-J_y^2$ and the previously reported CEF parameters for PrIr$_2$Zn$_{20}$; $B_4^0 = -0.0109$ K, $B_4^4$ = $5B_4^0$, $B_6^0 = -0.4477$ mK, and $B_6^4 = -21B_6^0$ (equivalent to: $W = -1.219$ K and $X = +0.537$ ~\cite{29LLW,17Iwasa13}). In the present setup, the rotationally invariant magneto-elastic effect~\cite{Dohm} can be neglected. Note that a tiny tetragonal distortion of $B_2^0 = +0.003$ K is added for the present calculation to reproduce the local minimum appearing at $\sim0.7$ T in the magnetic-field dependence of $(C_{11}-C_{12})/2$ at 0.04 K (arrowhead in the inset of Fig. 2). This may be caused by small inhomogeneities leading to a tiny disorder. This splits the $\Gamma_3$(E) doublet into two singlets with a gap of $\Delta_{\Gamma3} = 0.048$ K, which could also be responsible for the deviation from the $+\ln{T}$ dependence below $\sim0.06$ K at $0.5-1.5$ T, as shown below in Fig. 4. This small gap obeys Zeeman splitting in magnetic field and a recovery of the degeneracy should occur in the vicinity of 0.6 T for $H \parallel$ [001]. From these facts, it can be concluded that the present sample has a finite volume fraction of disordered Pr sites where the 4$f$ electrons are localized even in the NFL region. We also tested a random two-level (RTL) model~\cite{20Yamane18} with a normal distribution of the gap value by assuming $0 \leq \Delta_{\Gamma3} \leq 0.096$ K ($0 \leq B_2^0 \leq +0.006$ K) with a mean value of $B_2^0 = +0.003$ K. This RTL model does not reproduce the $+\ln{T}$ dependence of the elastic constant and further reveals the same result as the calculation done for a single mean value as shown by the solid curves in Fig. 3.

We can also rule out the possibilities of the contributions from the nuclear spin~\cite{30Aoki11, 31Kondo61} and the effect of quantum tunneling of atoms (so-called off-center degrees of freedom~\cite{24Goto04}) by following reasons. The dotted curves in Fig. 3 are the calculations including the additional contribution from hyperfine interactions of nuclear dipoles (rank 1) as described in the supplemental material. The nuclear contribution causes a minor deviation below $\sim0.1$ K in the present analysis and makes the fit even worse. On the other hand, the effect of quantum tunneling of atoms can simply be estimated from the 14-T data, since it can be considered that the off-center degrees of freedom must be magnetically robust. The background of the elastic constant $C_{\Gamma3}^0(T)$ used in the present analysis is estimated from the fit to the whole temperature range of the data at 14 T (also see details in the supplemental material). Indeed, the contribution from the off-center degrees of freedom is relatively small compared to the change of the $+\ln{T}$ softening. According to the above considerations, we conclude that both of the nuclear dipolar (hyperfine) interaction and the off-center tunneling phenomena cannot reproduce the $+\ln{T}$ behavior in the elastic constant.

\begin{figure}[t]
\includegraphics[width=0.85\linewidth]{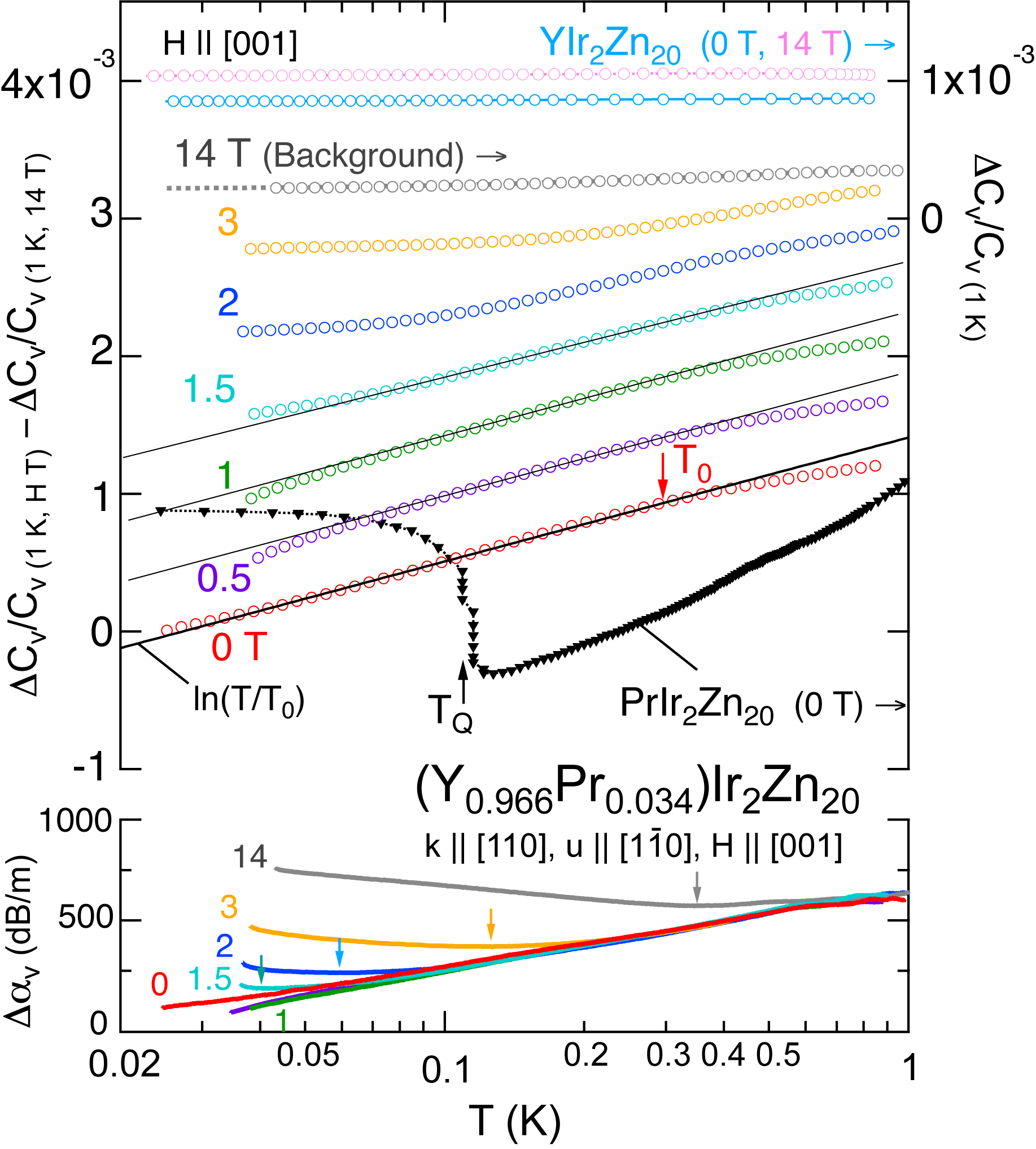}
\caption{\label{fig:fig4} (Upper panel) Temperature dependence of the background-subtracted elastic constant $(C_{11}-C_{12})/2$, shown as relative change, at selected magnetic fields. Data of PrIr$_2$Zn$_{20}$ ($x$ = 1)~\cite{22Ishii11} and YIr$_2$Zn$_{20}$ ($x$ = 0) are also displayed for comparison. The data are vertically shifted to facilitate display of the temperature and magnetic field variations.  (Lower panel) Ultrasonic attenuation coefficient vs. temperature.}
\end{figure}

In Fig. 4, we replot the data on a logarithmic $T$ scale in which the 14-T data, representing the background of the off-center contribution, has been subtracted. The background contribution is negligibly small as shown the 14-T data and $x$ = 0 (YIr$_2$Zn$_{20}$) system for comparison. It becomes obvious that the elastic constant $(C_{11}-C_{12})/2$ exhibits a $+\ln{(T/T_0)}$ dependence for temperature below $T_0 \sim0.3$ K and below 1.5 T. This logarithmic divergence of the quadrupolar susceptibility, {\it i.e.,} logarithmic softening of the elastic constant, is consistent with the prediction from the quadrupolar Kondo theory~\cite{Cox88, 7Cox98}, and also clearly identifies that the channel number of the Kondo effect in the present system must be two, in contrast to the single-channel Kondo effect (magnetic dipolar in origin), which leads to a leveling-off feature of the quadrupolar susceptibility ~\cite{Sacramento91,Kusunose16}. The small deviation below about 0.06 K ($\Delta_{\Gamma3}$) has been discussed above and is probably caused by tiny disorder. When applying magnetic field, the data deviates from the $+\ln{(T/T_0)}$ behavior and seems to crossover to localized electronic states above 3 T. The fact is also consistent with the preconceived scenario that the Zeeman splitting of not only the $\Gamma_3$(E) doublet of the local $4f$ electrons, due to the mixing of the first excited $\Gamma_4$(T$_1$) CEF state, but also the splitting of the $\Gamma_8$ states of the conduction band in finite magnetic fields will destroy the $c$-$f$ hybridization, which was driving the QKE. The ultrasonic attenuation coefficient is displayed in the lower panel of Fig. 4. The upturn appearing towards low temperatures in the ultrasonic attenuation (indicated by down arrows) strongly suggests a mixing of the excited $\Gamma_4$(T$_1$) triplet state in applied magnetic fields, while there is no enhancement or anomaly in the ultrasonic attenuation at fields below 1.5 T, which also rules out dynamical effects, such as the vibronic state~\cite{Araki12}. Based on the above investigations, the logarithmic temperature dependence of the elastic constant ($C_{11}-C_{12}$)/2 is considered to be clearly and strongly relevant to the recent finding of the NFL behavior in the electrical resistivity and specific heat of Y$_{0.966}$Pr$_{0.034}$Ir$_2$Zn$_{20}$ in the diluted Pr systems in the same temperature and magnetic field region. Further investigations are needed to check a possible systematic change in the logarithmic behavior toward a wide range of Pr concentrations.

In conclusion, we have reported that the elastic constant ($C_{11}-C_{12}$)/2, related to the $\Gamma_3$(E)-symmetry quadrupolar susceptibility, of Y$_{0.966}$Pr$_{0.034}$Ir$_2$Zn$_{20}$ obeys a Curie law down to $\sim1$ K. ($C_{11}-C_{12}$)/2 exhibits a strong deviation from the localized 4$f$-electron model and a logarithmic temperature dependence below $T_0 \sim0.3$ K, where NFL behavior in the specific heat and electrical resistivity have previously been reported. This logarithmic temperature variation manifested in the $\Gamma_3$(E)-symmetry quadrupolar susceptibility is consistent with the theoretical prediction of a QKE by Cox. Thus, it is reasonable to consider that the local NFL behavior of the present compound arises from a single-site two-channel (electric quadrupolar) Kondo effect.

\begin{acknowledgments}
The present research was supported by JSPS KAKENHI Grants Nos. JP15KK0169, JP18H04297, JP18H01182, JP17K05525, JP18KK0078, JP15KK0146, JP15H05882, JP15H05885, JP15H05886, JP15K21732, and the Strategic Young Researcher Overseas Visits Program for Accelerating Brain Circulation from JSPS. We acknowledge the support of the HLD at HZDR, a member of the European Magnetic Field Laboratory (EMFL). One of the authors T.Y. would like to thank J. Klotz for supporting the measurements at HZDR.
\end{acknowledgments}

\providecommand{\noopsort}[1]{}\providecommand{\singleletter}[1]{#1}%

\renewcommand{\thetable}{A.\arabic{table}}
\setcounter{table}{0}
\renewcommand{\thefigure}{A.\arabic{figure}}
\setcounter{figure}{0}
\onecolumngrid
\begin{center}
\vspace{5mm}
{\bf \large Supplemental material for\\ "Evidence for the Single-Site Quadrupolar Kondo Effect in the Dilute non-Kramers System Y$_{1-x}$Pr$_x$Ir$_2$Zn$_{20}$"}\\
\vspace{5mm}
T. Yanagisawa, H. Hidaka, H. Amitsuka, S. Zherlitsyn, J. Wosnitza, Y. Yamane, and T. Onimaru\\
\vspace{5mm}
\end{center}
\section{\label{sec:level1}1. Ultrasonic Methods}
Ultrasonic measurements have been widely used in studies of condensed matter. They are valuable tools for detailed characterization of the elasticity of lattice as well as any other type of excitations which couple to the strain fields with an appropriate symmetry induced by ultrasound in solids, {\it e.g.}, electric quadrupolar and hexadecapolar degrees of freedom, magnetic ordering (via magneto-elastic phenomena), superconductivity, local Einstein phonon, quantum oscillation (acoustic dHvA effect via electron-lattice coupling) and so on. By using longitudinal and transverse ultrasonic modes, the response of these excitations of electron and/or phonon systems can be obtained spectroscopically. In particular, the temperature dependence of the elastic constant is one of the powerful tools to investigate the non-Kramers $\Gamma_3$ ground state system, because the $\Gamma_3$-type elastic strain $\varepsilon_{\rm u}$  and $\varepsilon_{\rm v}$ corresponding to the transverse ultrasonic modes $C_{\rm u}$ and $C_{\rm v}$ couple to the the electric quadrupole with same symmetry as $O_u = \frac{1}{2}(2J_z^2-J_x^2-J_y^2)$ and $O_v = \frac{\sqrt{3}}{2}(J_x^2-J_y^2)$, respectively (Fig. A.1), which the non-magnetic $\Gamma_3$ ground doublet has. 
                                                                                                                                                                                                                                                                                                                                                                                                                                                                                                                                                                                                                             
\begin{figure}[h]
\includegraphics[width=0.5\linewidth]{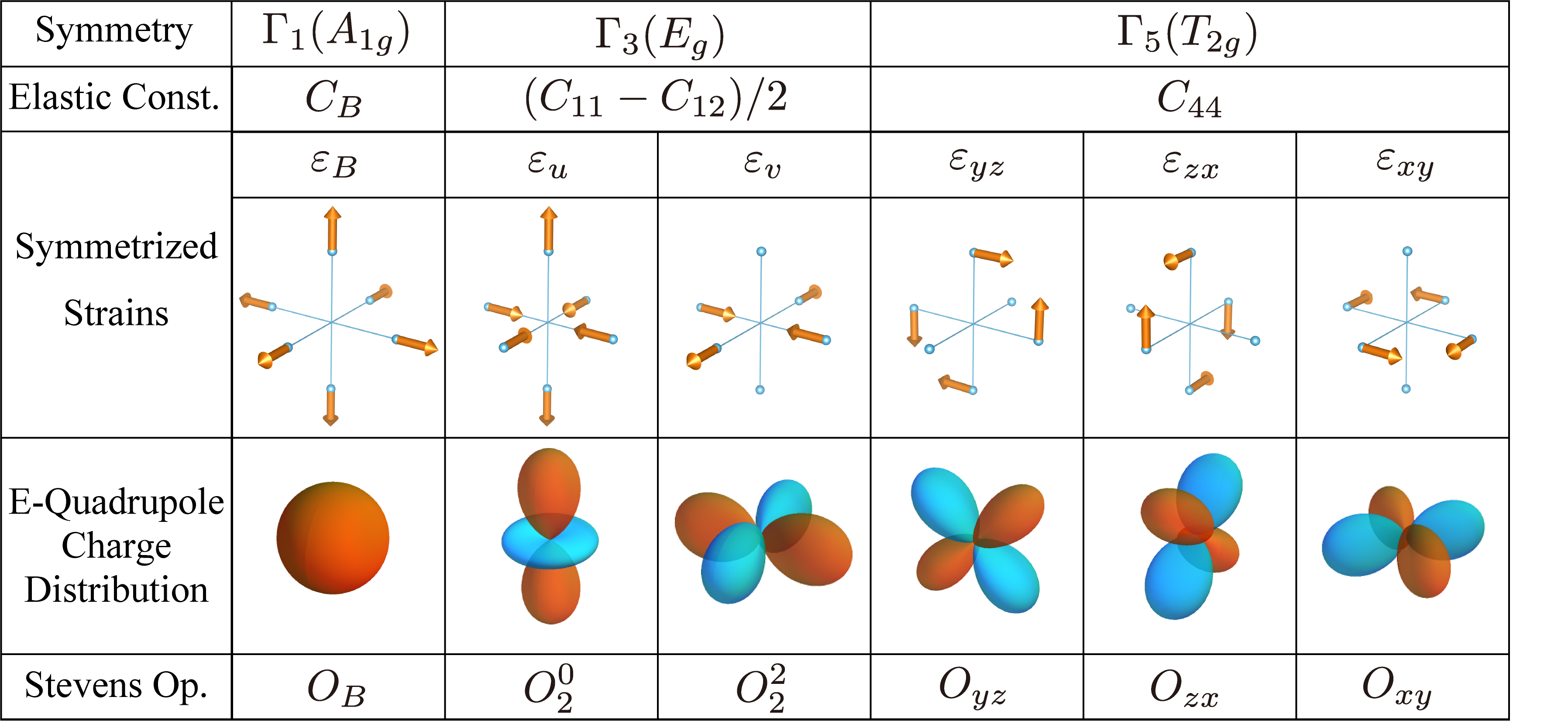}
\caption{\label{figure:fig1} }
Symmetry, elastic constant, symmetrized strain and quadrupole, and illustration of strain and coupled charge distributions.
\end{figure}

\begin{table}[h]
\caption{\label{tab:table1}Symmetry, strain, quadrupole and related elastic stuffiness constant}
{\begin{tabular}{@{}cccc}\toprule
Symmetry $\Gamma$			&Strain $\varepsilon_\Gamma$			&Quadrupole $O_\Gamma$&Elastic Stiffnes Constant $C_\Gamma$\\
\colrule
$\Gamma_{\rm 1}$ (A$_{\rm 1g})$ 	& $\varepsilon_{\rm B}=\varepsilon_{xx}-\varepsilon_{yy}-\varepsilon_{zz}$ 			&$\frac{1}{3}(J_x^2+J_y^2+J_z^2)$& $C_{\rm B} = C_{11}-\tfrac{4}{3}(C_{11}-C_{12})$\\
\colrule
$\Gamma_{\rm 3}$ (E$_{\rm g})$ 	& $\varepsilon_{\rm u}=(2\varepsilon_{zz}-\varepsilon_{xx}-\varepsilon_{yy})/\sqrt{3}$	&$\frac{1}{2}(2J_z^2-J_x^2-J_y^2)$& $C_{\rm u}=(C_{11}-C_{12})/2$\\
& $\varepsilon_{\rm v}=\varepsilon_{xx}-\epsilon_{yy}$										&$\frac{\sqrt{3}}{2}(J_x^2-J_y^2)$& $C_{\rm v}=(C_{11}-C_{12})/2$\\
\colrule
$\Gamma_{\rm 5}$ (T$_{\rm 2g})$	& $\varepsilon_{yz}$&$\frac{\sqrt{3}}{2}(J_yJ_z-J_zJ_y)$& $C_{44}$\\
& $\varepsilon_{zx}$												&$\frac{\sqrt{3}}{2}(J_zJ_x-J_xJ_z)$& $C_{44}$\\
& $\varepsilon_{xy}$										&$\frac{\sqrt{3}}{2}(J_xJ_y-J_yJ_x)$& $C_{44}$\\
\botrule
\end{tabular}}
\label{symbols}
\end{table}

\section{\label{sec:level1}2. Calculation of Quadrupolar Susceptibility under Cubic CEF}
Calculation of the temperature and magnetic field variations of the elastic constant in the present paper is performed by using the theory based on Wigner-Brillouin perturbation method. The crystalline electric field (CEF) Hamiltonian with elastic-strain mediated perturbation is,
\begin{equation}
 \mathscr{H}=\mathscr{H}_{\rm CEF}+\sum_{\Gamma}\frac{\partial \mathscr{H}_{\rm CEF}}{\partial \epsilon_\Gamma}\epsilon_\Gamma.
 \end{equation}
Here, $\epsilon_\Gamma$ is symmetrized strain with the point group symmetry $\Gamma$, which is induced by ultrasound. The Cubic CEF Hamiltonian taking Zeeman effect and tiny tetragonal distortion into consideration is written as
\begin{eqnarray}
 \mathscr{H}_{\rm CEF}&=&\mathscr{H}_{\rm Cubic}+\mathscr{H}_{\rm Zeeman}+\mathscr{H}_{\rm Tetra.}\\ \nonumber
&=&B_4^0(O_4^0+5O_4^4)+B_6^0(O_6^0-21O_6^4)+g_J \mu_{\rm B} \sum_{i=x,y,z}J_iH_i+B_2^0O_2^0.
 \end{eqnarray}
Here, $B_m^n$ are the CEF parameters and $O_m^n$ are the Stevens operators.
The second term of Eq. (1) is explained in terms of electric quadrupole-strain interaction written as
\begin{eqnarray}
\mathscr{H}_{\rm QS (\Gamma)}=-g_{\Gamma}O_{\Gamma}\epsilon_{\Gamma},
\end{eqnarray}
where $g_{\Gamma}$ is a coupling constant, and $O_{\Gamma}$ is an electric quadrupolar moment (as listed in Fig. 1).\\

The Helmholtz free energy of the local $4f$ electronic states in the CEF can be written as,
\begin{eqnarray}
F=U-Nk_{\rm B}T\ln \sum_{i}\exp\{-E_i(\epsilon_\Gamma)/k_{\rm B}T\},
\end{eqnarray}
where $N$ is the number of ions in a unit volume, $i$ is the number index for $J$ multiplets and their degenerate states. $U$ gives the internal energy for the strained system, which is written in terms of the symmetry strains and elastic constants,
\begin{eqnarray}
U=\frac{1}{2}\sum_{\Gamma}C_{\Gamma}\epsilon_{\Gamma}^2.
\end{eqnarray}

$E_i$($\epsilon_{\Gamma}$) is a perturbated CEF level as a function of strain $\epsilon_{\Gamma}$ up to the second-order perturbation, which can be considered as

\begin{eqnarray}
E_i(\epsilon_\Gamma)=E_i^0+g_\Gamma\left<i|\mathscr{H}_{\rm QS (\Gamma)}|i\right>\epsilon_{\Gamma}+g_\Gamma^2\sum_{j \neq i}\frac{|\left<j|\mathscr{H}_{\rm QS (\Gamma)}|i\right>|^2}{E_j^{(0)}-E_i^{(0)}}\epsilon_{\Gamma}^2.
\end{eqnarray}

In the second perturbation, the temperature dependence of the elastic constant is given by
\begin{equation}
C_\Gamma(T,H)=C_\Gamma^0-Ng_\Gamma^2\chi_\Gamma(T, H).
\end{equation}

Here, $C_\Gamma^0$ is the background of the elastic constant. The single-ion quadrupolar susceptibility $\chi_\Gamma$ is defined as the second derivative of the free energy with respect to strain (in the $\epsilon_\Gamma \rightarrow 0$ limit),
\begin{eqnarray}
-g_\Gamma^2\chi_\Gamma&=-\left<\frac{\partial^2E_i}{\partial\epsilon_\Gamma^2}\right>
+\frac{1}{k_{\rm B}T}\Bigl[\left<\Bigl(\frac{\partial E_i}{\partial\epsilon_\Gamma}\Bigr)^2 \right>-\left< \frac{\partial E_i}{\partial\epsilon_\Gamma}\right>^2\Bigr].
\end{eqnarray}

In addition to the strain-quadrupole interaction (Eq. 3), the intersite quadrupole-quadrupole interaction can also be added by using the molecular field approximation of quadrupolar moment $O_{\Gamma}$ by considering sublattices $\alpha$, as
\begin{eqnarray}
\mathscr{H}_{\rm QQ (\Gamma)}=-\sum_{\alpha}\sum_{\Gamma}g'_{\Gamma}\left<O_{\Gamma}\right>O_{\Gamma}^{\alpha}.
\end{eqnarray}

The temperature dependence of the elastic constant (Eq. 7) will be rewritten as following equation,
\begin{equation}
C_\Gamma(T,H)=C_\Gamma^0-\frac{Ng_\Gamma^2\chi_\Gamma(T, H)}{1-g'_\Gamma\chi_\Gamma(T, H)}.
\end{equation}

\section{\label{sec:level2}3. Nuclear Contribution}

In the present analysis, we also consider an additional term in the CEF hamiltonian, which is the hyperfine interaction between the nuclear dipole and $\Gamma_3$(E) electric quadrupole of the 4$f$ electrons. The dotted curves in Fig. 3 of the main text are the calculations including the additional contribution from hyperfine interactions of nuclear dipoles (rank 1) as described by following Hamiltonian,
\begin{equation}
\mathscr{H}_{\rm HF} = A_{\rm HF} I \cdot J -g_N\mu_N I\cdot\mu_0 H.
\end{equation}
Here, $I$ is the nuclear spin $I = 5/2$ of the $^{141}$Pr nucleus (natural abundance of 100\%) and $J$ is the total angular momentum $J = 4$ for Pr$^{3+}$. $g_N = 1.72$ and $\mu_N$ are the nuclear $g$-factor and the nuclear magneton, respectively. We used $A_{\rm hf} = +0.052$ K for the Pr nuclei. The nuclear contribution causes a minor deviation below $\sim0.1$ K in the present analysis. Note that a possible contribution from the nuclear quadrupoles (rank 2) is not considered here and should be tested in the future. The nuclear magnetic moment of the “off-cite” nucleus; $^{67}$Zn ($I$ = 3/2, natural abundance of 4.04\%), $^{89}$Y ($I$ = 1/2, natural abundance of 100\%), $^{193}$Ir ($I$ = 3/2, natural abundance of 62.7\%), $^{191}$Ir ($I$ = 3/2, natural abundance of 37.3\%), are negligible, since the “on-site” $^{141}$Pr hyperfine interaction should be dominant for the present analysis of the Pr’s single-ion susceptibility, compare to the weak dipolar fields from the “off-site” nucleus. In addition, it should be noted that the upturn of the specific heat appearing below 0.5 K in magnetic field of $B \le 12$ T is well reproduced by only considering Pr’s nuclear dipolar contribution. [Yamane {\it et al.}, AIP Advances {\bf 8}, 101338 (2018).]

\section{\label{sec:level3}4. Contribution from quantum tunneling of atoms\\
(Off-center degrees of freedom)}

Quantum tunneling (or off-center tunneling) is a local Einstein-phonon-like quantum oscillation of the atom through the potential hills between potential minima at around the high-symmetrical sites. This phenomenon has generally been found in glasses (as a two-level system) and some cage-structured compounds such as clathrate metals, filled-skutterudite compounds, and the present $RT_2$Zn$_{20}$ ($R$= Y, La or Pr, $T$ = Rh or Ir) compounds. The symmetrical, off-center mode couples to the appropriate symmetrized strain (deformation potential) induced by ultrasound, and causes magnetically-insensitive Curie-type softening of the elastic constant at low temperatures.\\

The effect of the quantum tunneling on the elastic constant can be considered as follows. The background $C_v^0(T)$ used in the present analysis includes a general phenomenological expressions of the phonon background and a Curie-type softening due to quantum tunneling which has been found in cage-structured compounds written as 
\begin{equation}
C_v^0(T) = a-b/\{\exp(t/c)-1\} +d(T -T_{\rm C})/(T-\Theta).
\end{equation}
Here, $a = 5.2771\times10^{10}$ J/m$^3$; $b = 0.005\times10^{10}$ J/m$^3$; $c$ = 20 K, $d = 0.0011\times10^{10}$ J/m$^3$ are used for the present analysis. The structural transition temperature (if the transition were second order) $T_{\rm C} = -0.6$ K and the transition temperature due to the two-ion (quadrupolar) interactions $\Theta = -0.3$ K are obtained from the fit to the whole temperature range of the data at 14 T, where the dHvA oscillations amplitude have been subtracted beforehand. The negative values of these parameters indicate an absence of the transitions. Indeed, the non-4$f$ system YIr$_2$Zn$_{20}$ also exhibits similar Curie-type softening in $(C_{11}-C_{12})/2$ mode, which evidences the presence of off-center contribution in the present system. Here, the softening of $0.002\%$ below 1 K in YIr$_2$Zn$_{20}$ is relatively smaller than that of $0.014\%$ in Y$_{0.966}$Pr$_{0.034}$Ir$_2$Zn$_{20}$ at 14 T as compared in Fig. 4 of the main text. However, the contribution from the off-center degrees of freedom is still negligibly small compared to the change of the $+\ln{T}$ softening as can be seen in Fig. 4 of the main text.

\section{\label{sec:level3}5. Estimation of the contribution from de Haas-van Alphen oscillation}

Here, we show how the background subtraction was done, regarding the temperature and magnetic field dependence of the dHvA oscillation. A constant background is used for the temperature dependence of the elastic constant $(C_{11}-C_{12})/2$ below 1 K, in order to represent the CEF+QKE contributions with absolute value of elastic constant in Fig. 3 (of the main text), and also to perform CEF analyses by considering volume fraction of Pr-ion and quadrupolar interactions. The schematic illustration of the background subtraction is shown in Fig. A.2. The value of the constant background is defined by the dHvA oscillation amplitude at the lowest temperature in each magnetic field. Figs. A.3(a, b), as shown below, represents the raw magnetic field dependences of the elastic constant $(C_{11}-C_{12})/2$. For comparison, the relative change version of the graph is also displayed in Figs. A.3(c, d). Here the data in Figs. A.3 (c, d) are normalized at 0 T and shifted with constant background subtraction by sorting the node of the waveform in order to clearly visualize the temperature variation of the CEF (Zeeman)+QKE effect and also dHvA oscillation amplitude under a magnetic field. The constant background of dHvA contributions in each magnetic field, for example $C_{\Gamma 3}^{\rm dHvA} = +0.005$ J/m$^3$ for 14 T, is defined from the dHvA oscillation amplitude at the lowest temperature (40 mK). In Fig. A.3(d), the temperature variation of the dHvA oscillation amplitude is visualized by zooming up at around 13.5 T. One can observe the change in the dHvA oscillation amplitude $\Delta C_v/C_v = 1.4\times 10^{-4}$ for 13.5 T below $\sim 1$ K is much smaller than the total amplitude $\Delta C_v/C_v \sim 1\times 10^{-3}$. The background subtraction is more important for the analysis of $+\ln T$ dependence in the low-magnetic field region. However, the change in the dHvA oscillation amplitude,
$\Delta C_v/C_v \sim 2.0\times 10^{-5}$ for 3 T, is two orders of magnitude less than the change in $+\ln T$ dependence of $\Delta C_v/C_v \sim 1.2\times 10^{-3}$ at 0 T. Therefore, it can be considered that the temperature dependence of the dHvA contribution below 1 K is negligible and the present constant background subtraction does not affect the conclusion presented in Fig. 4 (in the manuscript).

\begin{figure}[h]
\includegraphics[width=0.5\linewidth]{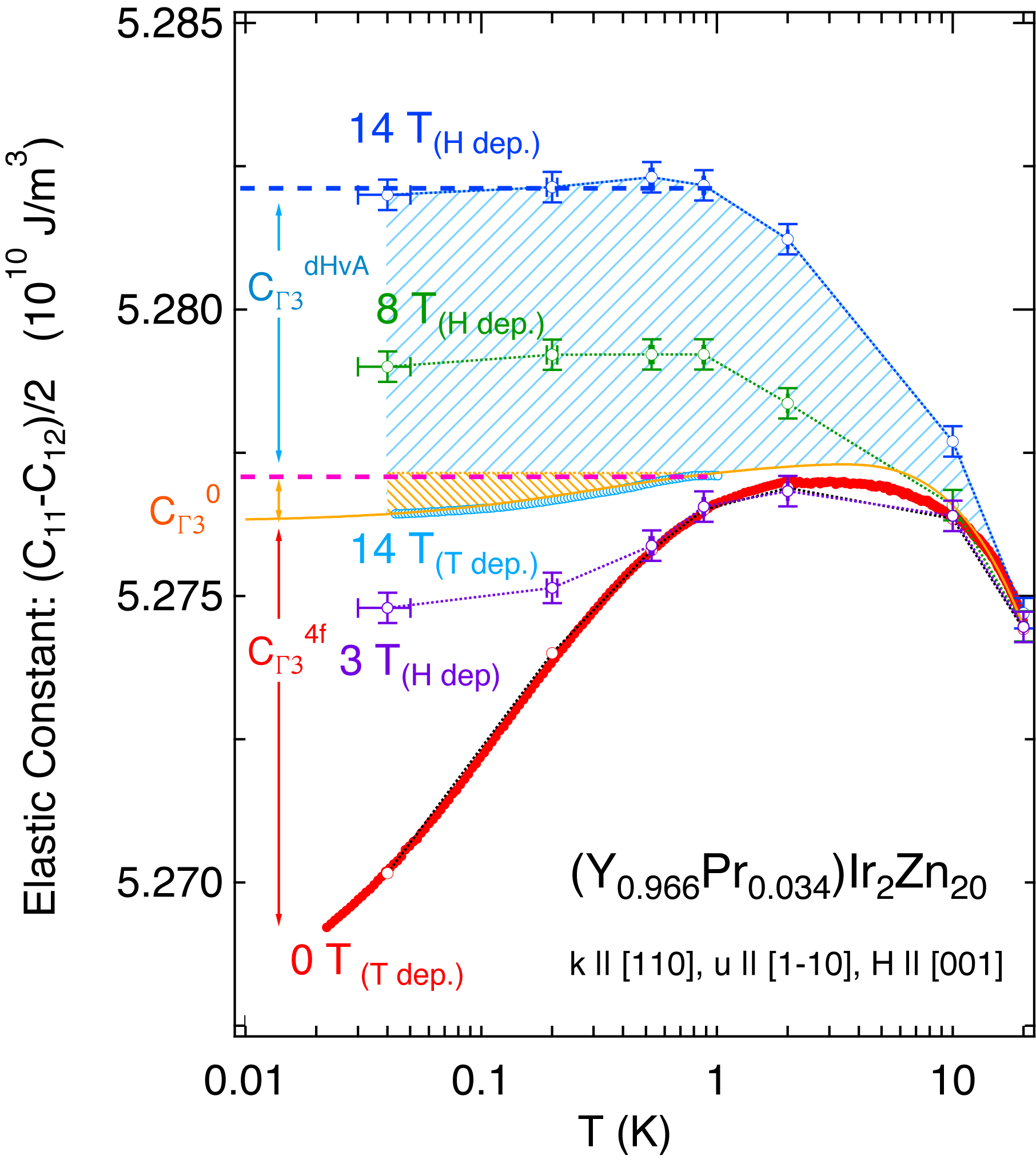}
\caption{\label{fig:figA} }
Schematic illustration of the background subtraction for the elastic constant $(C_{11}-C_{12})/2$ of Y$_{0.0966}$Pr$_{0.034}$Ir$_2$Zn$_{20}$ Open circles with error bars are the temperature dependence of the elastic constant, which were converted from the magnetic field dependence at fixed temperatures as shown in Fig. A.3.(a). The blue and orange hatched area indicate the estimated dHvA contributions $C_{\Gamma 3}^{\rm dHvA}$ and phonon contributions (including quantum tunneling) $C_{\Gamma 3}^{\rm 0}$ as the background for the CEF analysis, respectively.
\end{figure}

\begin{figure}[h]
\includegraphics[width=0.8\linewidth]{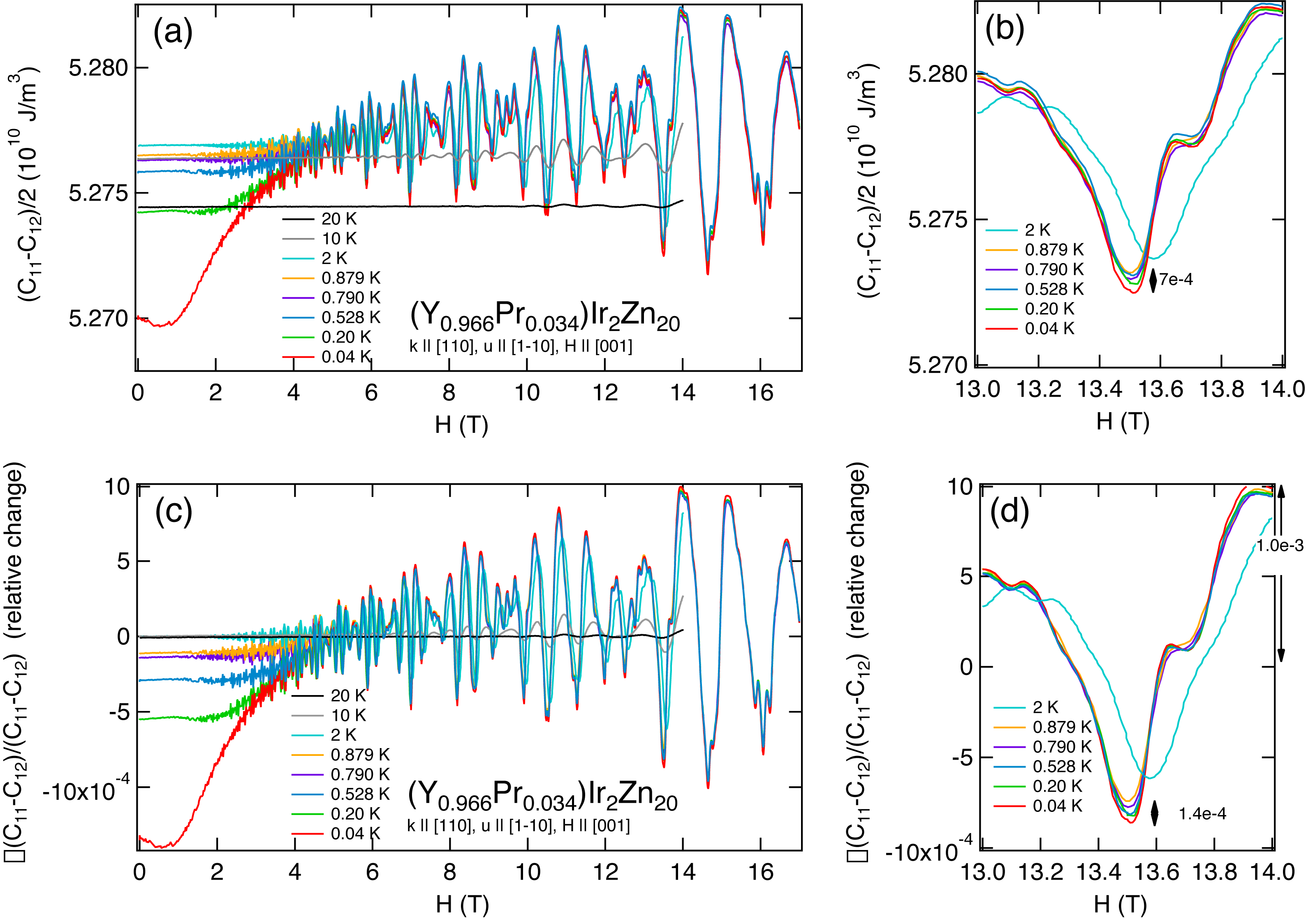}
\caption{\label{fig:figA} }
Magnetic field dependence of the elastic constant $(C_{11}-C_{12})/2$ of Y$_{0.0966}$Pr$_{0.034}$Ir$_2$Zn$_{20}$: (a) plot with absolute value without background subtraction, (c) plot with relative change, vertically shifted for easy comparison of the temperature dependence of the dHvA signal amplitudes, (b and d) zooming up around 13.5 T.
\end{figure}

\section{\label{sec:level3}6. Frequency Dependence}

In the following Fig. A4, we show the frequency dependence of the elastic constant $(C_{11}-C_{12})/2$ vs. temperature between 17.5 to 237 MHz. There is frequency dependent deviation below 10 K, which could be due to ultrasonic dispersion caused by a rattling effect, but the low-temperature $+\ln T$ dependence is not as much affected by frequency change (except for the data at 17.5 and 151 MHz, which have poor reproducibility due to relatively bad signal quality). The reason for the current choice of frequency 108 MHz in the main text is mainly due to impedance matching of the LiNbO$_3$ transducers, and the best signal quality.

\begin{figure}[h]
\includegraphics[width=0.6\linewidth]{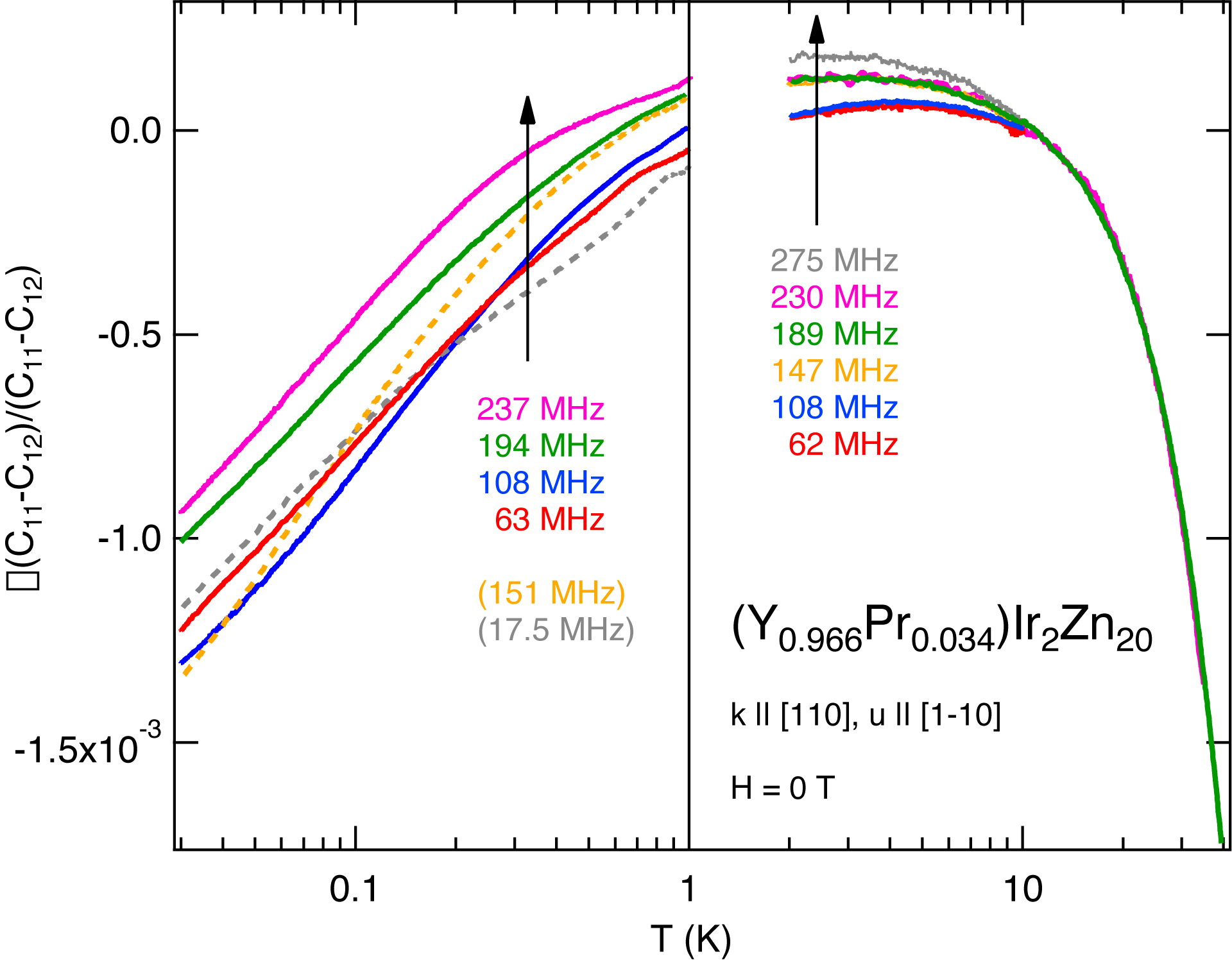}
\caption{\label{fig:figA} }
Frequency dependence of the elastic constant $(C_{11}-C_{12})/2$ of Y$_{0.0966}$Pr$_{0.034}$Ir$_2$Zn$_{20}$vs. temperature at 0 T
\end{figure}

\end{document}